\documentclass[a4paper]{llncs}

\usepackage{amsfonts}
\usepackage{graphicx}
\usepackage{caption}
\usepackage{subcaption}
\usepackage{footnote}
\usepackage{enumerate}

\RequirePackage{graphics}
\usepackage{xcolor}
\usepackage{footnote}
\usepackage{xparse}
\usepackage{cite}
\usepackage{booktabs}

\renewcommand{\thesection}{\arabic{section}}
\usepackage[noend]{algpseudocode}
\usepackage{algorithm}

\algblock{ParFor}{EndParFor}
\algnewcommand\algorithmicparfor{\textbf{parfor}}
\algnewcommand\algorithmicpardo{\textbf{do}}
\algnewcommand\algorithmicendparfor{\textbf{end\ parfor}}
\algrenewtext{ParFor}[1]{\algorithmicparfor\ #1\ \algorithmicpardo}
\algrenewtext{EndParFor}{\algorithmicendparfor}
\algnewcommand\algorithmicinput{\textbf{Input:}}
\algnewcommand\algorithmicoutput{\textbf{Output:}}
\algnewcommand\Input{\item[\algorithmicinput]}%
\algnewcommand\Output{\item[\algorithmicoutput]}%

\usepackage[hyphens]{url} 
\usepackage{amssymb}
\usepackage{amsmath}
\usepackage{keyval}
\usepackage{xspace}
\usepackage{paralist}
\usepackage{listings}
\usepackage{multirow}
\usepackage{stmaryrd}
\usepackage{adjustbox} 
\usepackage{makecell}
\usepackage{sidecap}
\usepackage{booktabs} 
\usepackage{threeparttable, tablefootnote}

\RequirePackage{tikz}
\usetikzlibrary{arrows,automata,shapes,calc,through,decorations.pathmorphing,decorations.fractals,chains,shapes.multipart}

\usepackage{arydshln} 
\usepackage[free-standing-units]{siunitx}
\usepackage{circuitikz}
\usepackage{tabularx} 

\def\titlename{NNV: The Neural Network Verification Tool for Deep Neural Networks and Learning-Enabled Cyber-Physical Systems\xspace}
\def\authortran{Hoang-Dung Tran}
\def\authorluan{Luan Viet Nguyen}
\def\authorweiming{Weiming Xiang}
\def\authorstan{Stanley Bak}
\def\authordiago{Diego Manzanas Lopez}
\def\authorpatrick{Patrick Musau}
\def\authorxiaodong{Xiaodong Yang}
\def\authortaylor{Taylor T. Johnson}

\usepackage[pdftex]{hyperref}
\hypersetup{
  colorlinks				=true,
  citecolor					={purple},
  linkcolor 				={blue},
  bookmarksnumbered	=true
}

\newcommand{\commenttaylor}[1]{}

\newcommand{\nnnum}[1]{\relax\ifmmode
  {\mathbb #1}_{\geq 0} \else ${\mathbb #1}_{\geq 0}$
  \fi}
\newcommand{\npnum}[1]{\relax\ifmmode
  {\mathbb #1}_{\leq 0} \else ${\mathbb #1}_{\leq 0}$
  \fi}
\newcommand{\pnum}[1]{\relax\ifmmode
  {\mathbb #1}_{> 0} \else ${\mathbb #1}_{> 0}$
  \fi}
\newcommand{\nnum}[1]{\relax\ifmmode
  {\mathbb #1}_{< 0} \else ${\mathbb #1}_{< 0}$
  \fi}
\newcommand{\plnum}[1]{\relax\ifmmode
  {\mathbb #1}_{+} \else ${\mathbb #1}_{+}$
  \fi}
\newcommand{\nenum}[1]{\relax\ifmmode
  {\mathbb #1}_{-} \else ${\mathbb #1}_{-}$
  \fi}

\newcommand{\extb}[1]{\relax\ifmmode {\sf ExtBeh}_{#1} \else ${\sf ExtBeh}_{#1}$\fi}
\newcommand{\tdists}[1]{\relax\ifmmode {\sf Tdists}_{#1} \else ${\sf Tdists}_{#1}$\fi}

\newcommand{\exec}[1]{\relax\ifmmode {\sf Execs}_{#1} \else ${\sf Exec}_{#1}$\fi}
\newcommand{\execf}[1]{\relax\ifmmode {\sf Execs}^*_{#1} \else ${\sf Exec}^*_{#1}$\fi}
\newcommand{\execi}[1]{\relax\ifmmode {\sf Execs}^\omega_{#1} \else ${\sf Exec}^\omega_{#1}$\fi}

\newcommand{\ctrace}[1]{\relax\ifmmode {\sf Ctraces}_{#1} \else ${\sf Ctraces}_{#1}$\fi}

\newcommand{\trace}[1]{\relax\ifmmode {\sf Traces}_{#1} \else ${\sf Traces}_{#1}$\fi}
\newcommand{\tracef}[1]{\relax\ifmmode {\sf Traces}^*_{#1} \else ${\sf Traces}^*_{#1}$\fi}
\newcommand{\tracei}[1]{\relax\ifmmode {\sf Traces}^\omega_{#1} \else ${\sf Traces}^\omega_{#1}$\fi}

\newcommand{\frag}[1]{\relax\ifmmode {\sf Frags}_{#1} \else ${\sf Frags}_{#1}$\fi}
\newcommand{\fragf}[1]{\relax\ifmmode {\sf Frags}^*_{#1} \else ${\sf Frags}^*_{#1}$\fi}
\newcommand{\fragi}[1]{\relax\ifmmode {\sf Frags}^\omega_{#1} \else ${\sf Frags}^\omega_{#1}$\fi}

\newcommand{\reach}[1]{\relax\ifmmode {\sf Reach}_{#1} \else ${\sf Reach}_{#1}$\fi}

\def\A{{\cal A}} 
\def\E{{\cal E}} 
\def\I{{\cal I}} 
\def\P{{\cal P}} 
\def\R{{\cal R}} 
\def\S{{\cal S}} 
\def\T{{\cal T}} 
\def\U{{\cal U}} 

\newcommand{\col}[1]{\relax\ifmmode \mathscr #1\else $\mathscr #1$\fi}

\definecolor{HIOAcolor}{rgb}{0.776,0.22,0.07}

\newcommand{\SC}[2]{\relax\ifmmode {\tt Scount}(#1,#2) \else ${\tt Scount}(#1,#2)$\fi}
\newcommand{\SCM}[2]{\relax\ifmmode {\tt Smin}(#1,#2) \else ${\tt Smin}(#1,#2)$\fi}
\newcommand{\Aut}[1]{\relax\ifmmode {\tt Aut}(#1) \else ${\tt Aut}(#1)$\fi}

\newcommand{\act}[1]{{\operatorname{\mathsf{#1}}}}

\renewcommand{\eqref}[1]{Equation~\ref{eq:#1}}

\newcommand{\remove}[1]{}
\newcommand{\salg}[1]{\relax\ifmmode {\mathcal F}_{#1}\else ${\mathcal F}_{#1}$\fi}
\newcommand{\msp}[1]{\relax\ifmmode (#1, \salg{#1}) \else $(#1, \salg{#1})$\fi}
\newcommand{\msprod}[2]{\relax\ifmmode ( #1 \times #2, \salg{#1} \otimes \salg{#2}) \else $(#1 \times #2, \salg{#1} \otimes \salg{#2})$\fi}
\newcommand{\dist}[1]{\relax\ifmmode {\mathcal P}\msp{#1}
  \else ${\mathcal P}\msp{#1}$\fi}
\newcommand{\subdist}[1]{\relax\ifmmode {\mathcal S}{\mathcal P}\msp{#1}
  \else ${\mathcal S}{\mathcal P}\msp{#1}$\fi}
\newcommand{\disc}[1]{\relax\ifmmode {\sf Disc}(#1)
  \else ${\sf Disc}(#1)$\fi}

\newcommand{\Trajeq}{\relax\ifmmode {\mathcal R}_\T \else ${\mathcal R}_\T$\fi}
\newcommand{\Acteq}{\relax\ifmmode {\mathcal R}_A \else ${\mathcal R}_A$\fi}
\newcommand{\noop}{\relax\ifmmode \lambda \else $\lambda$\fi}
\newcommand{\close}[1]{\relax\ifmmode \overline{#1} \else $\overline{#1}$\fi}

\newcommand{\tup}[1]
           {
             \relax\ifmmode
             \langle #1 \rangle
             \else $\langle$ #1 $\rangle$ \fi
           }

\newcommand{\lit}[1]{ \relax\ifmmode
                \mathord{\mathcode`\-="702D\sf #1\mathcode`\-="2200}
                \else {\it #1} \fi }

\newcommand{\figuresize}{\scriptsize}

\lstdefinelanguage{ioa}{
  basicstyle=\figuresize,
  keywordstyle=\bf \figuresize,
  identifierstyle=\it \figuresize,
  emphstyle=\tt \figuresize,
  mathescape=true,
  tabsize=20,
  sensitive=false,
  columns=fullflexible,
  keepspaces=false,
  flexiblecolumns=true,
  basewidth=0.05em,
  escapeinside={(*@}{@*)},
  moredelim=[il][\rm]{//},
  moredelim=[is][\sf \figuresize]{!}{!},
  moredelim=[is][\bf \figuresize]{*}{*},
  keywords={automaton,and,
  	 choose,const,continue, components,
  	 discrete, do,
  	 eff, Eff, external,else, elseif, evolve, end,
  	 fi,for, forward, from,
  	 hidden,
  	 in,input,internal,if,invariant, initially, imports,
     let,
     or, output, operators, od, of,
     pre, Pre,
     return,
     such,satisfies, stop, signature, simulation,
     trajectories,trajdef, transitions, that,then, type, types, to, tasks,
     variables, vocabulary,
     when,where, with,while},
  emph={set, seq, tuple, map, array, enumeration},
   literate=
        {(}{{$($}}1
        {)}{{$)$}}1
        {\\in}{{$\in\ $}}1
        {\\preceq}{{$\preceq\ $}}1
        {\\subset}{{$\subset\ $}}1
        {\\subseteq}{{$\subseteq\ $}}1
        {\\supset}{{$\supset\ $}}1
        {\\supseteq}{{$\supseteq\ $}}1
        {\\forall}{{$\forall$}}1
        {\\le}{{$\le\ $}}1
        {\\ge}{{$\ge\ $}}1
        {\\gets}{{$\gets\ $}}1
        {\\cup}{{$\cup\ $}}1
        {\\cap}{{$\cap\ $}}1
        {\\langle}{{$\langle$}}1
        {\\rangle}{{$\rangle$}}1
        {\\exists}{{$\exists\ $}}1
        {\\bot}{{$\bot$}}1
        {\\rip}{{$\rip$}}1
        {\\emptyset}{{$\emptyset$}}1
        {\\notin}{{$\notin\ $}}1
        {\\not\\exists}{{$\not\exists\ $}}1
        {\\ne}{{$\ne\ $}}1
        {\\to}{{$\to\ $}}1
        {\\implies}{{$\implies\ $}}1
        {<}{{$<\ $}}1
        {>}{{$>\ $}}1
        {=}{{$=\ $}}1
        {~}{{$\neg\ $}}1
        {|}{{$\mid$}}1
        {'}{{$^\prime$}}1
        {\\A}{{$\forall\ $}}1
        {\\E}{{$\exists\ $}}1
        {\\nE}{{$\nexists\ $}}1
        {\\/}{{$\vee\,$}}1
        {\\vee}{{$\vee\,$}}1
        {/\\}{{$\wedge\,$}}1
        {\\wedge}{{$\wedge\,$}}1
        {=>}{{$\Rightarrow\ $}}1
        {->}{{$\rightarrow\ $}}1
        {<=}{{$\Leftarrow\ $}}1
        {<-}{{$\leftarrow\ $}}1
        {~=}{{$\neq\ $}}1
        {\\U}{{$\cup\ $}}1
        {\\I}{{$\cap\ $}}1
        {|-}{{$\vdash\ $}}1
        {-|}{{$\dashv\ $}}1
        {<<}{{$\ll\ $}}2
        {>>}{{$\gg\ $}}2
        {||}{{$\|$}}1
        {[}{{$[$}}1
        {]}{{$\,]$}}1
        {[[}{{$\langle$}}1
        {]]]}{{$]\rangle$}}1
        {]]}{{$\rangle$}}1
        {<=>}{{$\Leftrightarrow\ $}}2
        {<->}{{$\leftrightarrow\ $}}2
        {(+)}{{$\oplus\ $}}1
        {(-)}{{$\ominus\ $}}1
        {_i}{{$_{i}$}}1
        {_j}{{$_{j}$}}1
        {_{i,j}}{{$_{i,j}$}}3
        {_{j,i}}{{$_{j,i}$}}3
        {_0}{{$_0$}}1
        {_1}{{$_1$}}1
        {_2}{{$_2$}}1
        {_n}{{$_n$}}1
        {_p}{{$_p$}}1
        {_k}{{$_n$}}1
        {-}{{$\ms{-}$}}1
        {@}{{}}0
        {\\delta}{{$\delta$}}1
        {\\R}{{$\R$}}1
        {\\Rplus}{{$\Rplus$}}1
        {\\N}{{$\N$}}1
        {\\times}{{$\times\ $}}1
        {\\tau}{{$\tau$}}1
        {\\alpha}{{$\alpha$}}1
        {\\beta}{{$\beta$}}1
        {\\gamma}{{$\gamma$}}1
        {\\ell}{{$\ell\ $}}1
        {--}{{$-\ $}}1
        {\\TT}{{\hspace{1.5em}}}3
      }

\lstdefinelanguage{ioaNums}[]{ioa}
{
  numbers=left,
  numberstyle=\tiny,
  stepnumber=2,
  numbersep=4pt
}

\lstdefinelanguage{ioaNumsRight}[]{ioa}
{
  numbers=right,
  numberstyle=\tiny,
  stepnumber=2,
  numbersep=4pt
}

\lstnewenvironment{IOA}%
  {\lstset{language=IOA}}
  {}

\lstnewenvironment{IOANums}%
  {
  \if@firstcolumn
    \lstset{language=IOA, numbers=left, firstnumber=auto}
  \else
    \lstset{language=IOA, numbers=right, firstnumber=auto}
  \fi
  }
  {}

\lstnewenvironment{IOANumsRight}%
  {
    \lstset{language=IOA, numbers=right, firstnumber=auto}
  }
  {}

\newcommand{\linefigioa}[9]{

}

\lstdefinelanguage{ioaLang}{%
  basicstyle=\ttfamily\small,
  keywordstyle=\rmfamily\bfseries\small,
  identifierstyle=\small,
  keywords={assumes,automaton,axioms,backward,bounds,by,case,choose,components,const,d,det,discrete,do,eff,else,elseif,ensuring,enumeration,evolve,fi,fire,follow,for,forward,from,hidden,if,in,%
    input,initially,internal,invariant,let, local,od,of,output,pre,schedule,signature,so,%
    simulation,states,variables, tasks, stop,tasks,that,then,to,trajdef,trajectory,trajectories,transitions,tuple,type,union,urgent,uses,when,where,while,yield},
  literate=
        {\\in}{{$\in$}}1
        {\\preceq}{{$\preceq$}}1
        {\\subset}{{$\subset$}}1
        {\\subseteq}{{$\subseteq$}}1
        {\\supset}{{$\supset$}}1
        {\\supseteq}{{$\supseteq$}}1
        {\\rho}{{$\rho$}}1
        {\\infty}{{$\infty$}}1
        {<}{{$<$}}1
        {>}{{$>$}}1
        {=}{{$=$}}1
        {~}{{$\neg$}}1
        {|}{{$\mid$}}1
        {'}{{$^\prime$}}1
        {\\A}{{$\forall$}}1 {\\E}{{$\exists$}}1
        {\\/}{{$\vee$}}1 {/\\}{{$\wedge$}}1
        {=>}{{$\Rightarrow$}}1
        {->}{{$\rightarrow$}}1
        {<=}{{$\leq$}}1 {>=}{{$\geq$}}1 {~=}{{$\neq$}}1
        {\\U}{{$\cup$}}1 {\\I}{{$\cap$}}1
        {|-}{{$\vdash$}}1 {-|}{{$\dashv$}}1
        {<<}{{$\ll$}}2 {>>}{{$\gg$}}2
        {||}{{$\|$}}1
        {<=>}{{$\Leftrightarrow$}}2
        {<->}{{$\leftrightarrow$}}2
        {(+)}{{$\oplus$}}1
        {(-)}{{$\ominus$}}1
}

\lstdefinelanguage{bigIOALang}{%
  basicstyle=\ttfamily,
  keywordstyle=\rmfamily\bfseries,
  identifierstyle=,
  keywords={assumes,automaton,axioms,backward,by,case,choose,components,const,%
    d,det,discrete,do,eff,else,elseif,ensuring,enumeration,evolve,fi,for,forward,from,hidden,if,in%
    input,initially,internal,invariant,local,od,of,output,pre,schedule,signature,so,%
    tasks, simulation,states,stop,tasks,that,then,to,trajdef,trajectories,transitions,tuple,type,union,urgent,uses,when,where,yield},
  literate=
        {\\in}{{$\in$}}1
        {\\preceq}{{$\preceq$}}1
        {\\subset}{{$\subset$}}1
        {\\subseteq}{{$\subseteq$}}1
        {\\supset}{{$\supset$}}1
        {\\supseteq}{{$\supseteq$}}1
        {<}{{$<$}}1
        {>}{{$>$}}1
        {=}{{$=$}}1
        {~}{{$\neg$}}1
        {|}{{$\mid$}}1
        {'}{{$^\prime$}}1
        {\\A}{{$\forall$}}1 {\\E}{{$\exists$}}1
        {\\/}{{$\vee$}}1 {/\\}{{$\wedge$}}1
        {=>}{{$\Rightarrow$}}1
        {->}{{$\rightarrow$}}1
        {<=}{{$\leq$}}1 {>=}{{$\geq$}}1 {~=}{{$\neq$}}1
        {\\U}{{$\cup$}}1 {\\I}{{$\cap$}}1
        {|-}{{$\vdash$}}1 {-|}{{$\dashv$}}1
        {<<}{{$\ll$}}2 {>>}{{$\gg$}}2
        {||}{{$\|$}}1
        {<=>}{{$\Leftrightarrow$}}2
        {<->}{{$\leftrightarrow$}}2
        {(+)}{{$\oplus$}}1
        {(-)}{{$\ominus$}}1
}

\lstnewenvironment{BigIOA}%
  {\lstset{language=bigIOALang,basicstyle=\ttfamily}
   \csname lst@SetFirstLabel\endcsname}
  {\csname lst@SaveFirstLabel\endcsname\vspace{-4pt}\noindent}

\lstnewenvironment{SmallIOA}%
  {\lstset{language=ioaLang,basicstyle=\ttfamily\scriptsize}
   \csname lst@SetFirstLabel\endcsname}
  {\csname lst@SaveFirstLabel\endcsname\noindent}

\newlength{\bracklen}

\renewcommand{\arraystretch}{\defaultArraystretch}

\newcommand{\tri}[3]{\ensuremath{\mathit{#1}^\mathit{#2}_\mathit{#3}}}

\newcommand{\sugLocalVars}[2]{\ifthenelse{\equal{}{#2}}%
                             {\tri{localVars}{#1}{desug}}%
                             {\tri{localVars}{#1}{#2,desug}}}
\newcommand{\sugVars}[2]{\ifthenelse{\equal{}{#2}}%
                        {\tri{vars}{#1}{desug}}%
                        {\tri{vars}{#1}{#2,desug}}}

\newenvironment{subSyntax}{\begin{array}{l}}{\end{array}}

\newcommand{\ms}[1]{\ifmmode%
\mathord{\mathcode`-="702D\it #1\mathcode`\-="2200}\else%
$\mathord{\mathcode`-="702D\it #1\mathcode`\-="2200}$\fi}

\def\A{{\cal A}} 
\def\T{{\cal T}} 

\lstdefinelanguage{pvs}{
  basicstyle=\tt \figuresize,
  keywordstyle=\sc \figuresize,
  identifierstyle=\it \figuresize,
  emphstyle=\tt \figuresize,
  mathescape=true,
  tabsize=20,
  sensitive=false,
  columns=fullflexible,
  keepspaces=false,
  flexiblecolumns=true,
  basewidth=0.05em,
  moredelim=[il][\rm]{//},
  moredelim=[is][\sf \figuresize]{!}{!},
  moredelim=[is][\bf \figuresize]{*}{*},
  keywords={and,
  	 begin,
  	 cases, const,
  	 do,
  	 external, else, exists, end, endcases, endif,
  	 fi,for, forall, from,
  	 hidden,
  	 in, if, importing,
     let, lambda, lemma,
     measure,
     not,
     or, of,
     return, recursive,
     stop,
     theory, that,then, type, types, type+, to, theorem,
     var,
     with,while},
  emph={nat, setof, sequence, eq, tuple, map, array, enumeration, bool, real, exp, nnreal, posreal},
   literate=
        {(}{{$($}}1
        {)}{{$)$}}1
        {\\in}{{$\in\ $}}1
        {\\mapsto}{{$\rightarrow\ $}}1
        {\\preceq}{{$\preceq\ $}}1
        {\\subset}{{$\subset\ $}}1
        {\\subseteq}{{$\subseteq\ $}}1
        {\\supset}{{$\supset\ $}}1
        {\\supseteq}{{$\supseteq\ $}}1
        {\\forall}{{$\forall$}}1
        {\\le}{{$\le\ $}}1
        {\\ge}{{$\ge\ $}}1
        {\\gets}{{$\gets\ $}}1
        {\\cup}{{$\cup\ $}}1
        {\\cap}{{$\cap\ $}}1
        {\\langle}{{$\langle$}}1
        {\\rangle}{{$\rangle$}}1
        {\\exists}{{$\exists\ $}}1
        {\\bot}{{$\bot$}}1
        {\\rip}{{$\rip$}}1
        {\\emptyset}{{$\emptyset$}}1
        {\\notin}{{$\notin\ $}}1
        {\\not\\exists}{{$\not\exists\ $}}1
        {\\ne}{{$\ne\ $}}1
        {\\to}{{$\to\ $}}1
        {\\implies}{{$\implies\ $}}1
        {<}{{$<\ $}}1
        {>}{{$>\ $}}1
        {=}{{$=\ $}}1
        {~}{{$\neg\ $}}1
        {|}{{$\mid$}}1
        {'}{{$^\prime$}}1
        {\\A}{{$\forall\ $}}1
        {\\E}{{$\exists\ $}}1
        {\\/}{{$\vee\,$}}1
        {\\vee}{{$\vee\,$}}1
        {/\\}{{$\wedge\,$}}1
        {\\wedge}{{$\wedge\,$}}1
        {->}{{$\rightarrow\ $}}1
        {=>}{{$\Rightarrow\ $}}1
        {->}{{$\rightarrow\ $}}1
        {<=}{{$\Leftarrow\ $}}1
        {<-}{{$\leftarrow\ $}}1
        {~=}{{$\neq\ $}}1
        {\\U}{{$\cup\ $}}1
        {\\I}{{$\cap\ $}}1
        {|-}{{$\vdash\ $}}1
        {-|}{{$\dashv\ $}}1
        {<<}{{$\ll\ $}}2
        {>>}{{$\gg\ $}}2
        {||}{{$\|$}}1
        {[}{{$[$}}1
        {]}{{$\,]$}}1
        {[[}{{$\langle$}}1
        {]]]}{{$]\rangle$}}1
        {]]}{{$\rangle$}}1
        {<=>}{{$\Leftrightarrow\ $}}2
        {<->}{{$\leftrightarrow\ $}}2
        {(+)}{{$\oplus\ $}}1
        {(-)}{{$\ominus\ $}}1
        {_i}{{$_{i}$}}1
        {_j}{{$_{j}$}}1
        {_{i,j}}{{$_{i,j}$}}3
        {_{j,i}}{{$_{j,i}$}}3
        {_0}{{$_0$}}1
        {_1}{{$_1$}}1
        {_2}{{$_2$}}1
        {_n}{{$_n$}}1
        {_p}{{$_p$}}1
        {_k}{{$_n$}}1
        {-}{{$\ms{-}$}}1
        {@}{{}}0
        {\\delta}{{$\delta$}}1
        {\\R}{{$\R$}}1
        {\\Rplus}{{$\Rplus$}}1
        {\\N}{{$\N$}}1
        {\\times}{{$\times\ $}}1
        {\\tau}{{$\tau$}}1
        {\\alpha}{{$\alpha$}}1
        {\\beta}{{$\beta$}}1
        {\\gamma}{{$\gamma$}}1
        {\\ell}{{$\ell\ $}}1
        {--}{{$-\ $}}1
        {\\TT}{{\hspace{1.5em}}}3
      }

\lstdefinelanguage{BigPVS}{
  basicstyle=\tt,
  keywordstyle=\sc,
  identifierstyle=\it,
  emphstyle=\tt ,
  mathescape=true,
  tabsize=20,
  sensitive=false,
  columns=fullflexible,
  keepspaces=false,
  flexiblecolumns=true,
  basewidth=0.05em,
  moredelim=[il][\rm]{//},
  moredelim=[is][\sf \figuresize]{!}{!},
  moredelim=[is][\bf \figuresize]{*}{*},
  keywords={and,
  	 begin,
  	 cases, const,
  	 do, datatype,
  	 external, else, exists, end, endif, endcases,
  	 fi,for, forall, from,
  	 hidden,
  	 in, if, importing,
     let, lambda, lemma,
     measure,
     not,
     or, of,
     return, recursive,
     stop,
     theory, that,then, type, types, type+, to, theorem,
     var,
     with,while},
  emph={nat, setof, sequence, eq, tuple, map, array, first, rest, add, enumeration, bool, real, posreal, nnreal},
   literate=
        {(}{{$($}}1
        {)}{{$)$}}1
        {\\in}{{$\in\ $}}1
        {\\mapsto}{{$\rightarrow\ $}}1
        {\\preceq}{{$\preceq\ $}}1
        {\\subset}{{$\subset\ $}}1
        {\\subseteq}{{$\subseteq\ $}}1
        {\\supset}{{$\supset\ $}}1
        {\\supseteq}{{$\supseteq\ $}}1
        {\\forall}{{$\forall$}}1
        {\\le}{{$\le\ $}}1
        {\\ge}{{$\ge\ $}}1
        {\\gets}{{$\gets\ $}}1
        {\\cup}{{$\cup\ $}}1
        {\\cap}{{$\cap\ $}}1
        {\\langle}{{$\langle$}}1
        {\\rangle}{{$\rangle$}}1
        {\\exists}{{$\exists\ $}}1
        {\\bot}{{$\bot$}}1
        {\\rip}{{$\rip$}}1
        {\\emptyset}{{$\emptyset$}}1
        {\\notin}{{$\notin\ $}}1
        {\\not\\exists}{{$\not\exists\ $}}1
        {\\ne}{{$\ne\ $}}1
        {\\to}{{$\to\ $}}1
        {\\implies}{{$\implies\ $}}1
        {<}{{$<\ $}}1
        {>}{{$>\ $}}1
        {=}{{$=\ $}}1
        {~}{{$\neg\ $}}1
        {|}{{$\mid$}}1
        {'}{{$^\prime$}}1
        {\\A}{{$\forall\ $}}1
        {\\E}{{$\exists\ $}}1
        {\\/}{{$\vee\,$}}1
        {\\vee}{{$\vee\,$}}1
        {/\\}{{$\wedge\,$}}1
        {\\wedge}{{$\wedge\,$}}1
        {->}{{$\rightarrow\ $}}1
        {=>}{{$\Rightarrow\ $}}1
        {->}{{$\rightarrow\ $}}1
        {<=}{{$\Leftarrow\ $}}1
        {<-}{{$\leftarrow\ $}}1
        {~=}{{$\neq\ $}}1
        {\\U}{{$\cup\ $}}1
        {\\I}{{$\cap\ $}}1
        {|-}{{$\vdash\ $}}1
        {-|}{{$\dashv\ $}}1
        {<<}{{$\ll\ $}}2
        {>>}{{$\gg\ $}}2
        {||}{{$\|$}}1
        {[}{{$[$}}1
        {]}{{$\,]$}}1
        {[[}{{$\langle$}}1
        {]]]}{{$]\rangle$}}1
        {]]}{{$\rangle$}}1
        {<=>}{{$\Leftrightarrow\ $}}2
        {<->}{{$\leftrightarrow\ $}}2
        {(+)}{{$\oplus\ $}}1
        {(-)}{{$\ominus\ $}}1
        {_i}{{$_{i}$}}1
        {_j}{{$_{j}$}}1
        {_{i,j}}{{$_{i,j}$}}3
        {_{j,i}}{{$_{j,i}$}}3
        {_0}{{$_0$}}1
        {_1}{{$_1$}}1
        {_2}{{$_2$}}1
        {_n}{{$_n$}}1
        {_p}{{$_p$}}1
        {_k}{{$_n$}}1
        {-}{{$\ms{-}$}}1
        {@}{{}}0
        {\\delta}{{$\delta$}}1
        {\\R}{{$\R$}}1
        {\\Rplus}{{$\Rplus$}}1
        {\\N}{{$\N$}}1
        {\\times}{{$\times\ $}}1
        {\\tau}{{$\tau$}}1
        {\\alpha}{{$\alpha$}}1
        {\\beta}{{$\beta$}}1
        {\\gamma}{{$\gamma$}}1
        {\\ell}{{$\ell\ $}}1
        {--}{{$-\ $}}1
        {\\TT}{{\hspace{1.5em}}}3
      }

\lstdefinelanguage{pvsNums}[]{pvs}
{
  numbers=left,
  numberstyle=\tiny,
  stepnumber=2,
  numbersep=4pt
}

\lstdefinelanguage{pvsNumsRight}[]{pvs}
{
  numbers=right,
  numberstyle=\tiny,
  stepnumber=2,
  numbersep=4pt
}

\lstnewenvironment{BigPVS}%
  {\lstset{language=BigPVS}}
  {}

\lstnewenvironment{PVSNums}%
  {
  \if@firstcolumn
    \lstset{language=pvs, numbers=left, firstnumber=auto}
  \else
    \lstset{language=pvs, numbers=right, firstnumber=auto}
  \fi
  }
  {}

\lstnewenvironment{PVSNumsRight}%
  {
    \lstset{language=pvs, numbers=right, firstnumber=auto}
  }
  {}

\newcommand{\linefigpvs}[9]{

}

\lstdefinelanguage{pvsproof}{
  basicstyle=\tt \figuresize,
  mathescape=true,
  tabsize=4,
  sensitive=false,
  columns=fullflexible,
  keepspaces=false,
  flexiblecolumns=true,
  basewidth=0.05em,
}

\def\N{\act{N}}

\newcommand{\localvar}[2]{{{#1_{#2}}}}

\def\xi{\localvar{x}{i}}

\def\reach{{\sf Reach}}

\def\Xi{\mathit{X_i}}

\begin{document}
%

\title{\titlename}

%
%
%

\author{\authortran \inst{1}, \authorxiaodong \inst{1}, \authordiago \inst{1}, \authorpatrick \inst{1}, \authorluan \inst{2}, \authorweiming \inst{1}, \authorstan \and \authortaylor \inst{1}}

\authorrunning{Tran et al.}
%
\institute{Institute for Software Integrated Systems, Vanderbilt University, TN, USA \and
  Department of Computer and Information Science, University of Pennsylvania, PA, USA}


%
%
\maketitle


\begin{abstract}
This paper presents the Neural Network Verification (NNV) software tool, a set-based verification framework for deep neural networks (DNNs) and learning-enabled cyber-physical systems (CPS). The crux of NNV is a collection of reachability algorithms that make use of a variety of set representations, such as polyhedra, star sets, zonotopes, and abstract-domain representations. NNV supports both exact (sound and complete) and over-approximate (sound) reachability algorithms for verifying safety and robustness properties of feed-forward neural networks (FFNNs) with various activation functions.  For learning-enabled CPS, such as closed-loop control systems incorporating neural networks, NNV provides exact and over-approximate reachability analysis schemes for linear plant models and FFNN controllers with piecewise-linear activation functions, such as ReLUs. For similar neural network control systems (NNCS) that instead have nonlinear plant models, NNV supports over-approximate analysis by combining the star set analysis used for FFNN controllers with zonotope-based analysis for nonlinear plant dynamics building on CORA. We evaluate NNV using two real-world case studies: the first is safety verification of ACAS Xu networks, and the second deals with the safety verification of a deep learning-based adaptive cruise control system.

%

\end{abstract}
\section{Introduction}
Deep neural networks (DNNs) have quickly become one of the most widely used tools for dealing with complex and challenging problems in numerous domains, such as image classification \cite{krizhevsky2012imagenet, cirecsan2012multi, gatys2016image}, function approximation, and natural language translation \cite{collobert2008unified, goldberg2016primer}.
Recently, DNNs have been used in safety-critical cyber-physical systems (CPS), such as autonomous vehicles \cite{wu2017squeezedet, chen2015deepdriving, bojarski2016end} and air traffic collision avoidance systems \cite{julian2016policy}.
Although utilizing DNNs in safety-critical applications can demonstrate considerable performance benefits, assuring the safety and robustness of these systems is challenging because DNNs possess complex non-linear characteristics.
Moreover, it has been demonstrated that their behavior can be unpredictable due to slight perturbations in their inputs (i.e., adversarial perturbations)~\cite{szegedy2013intriguing}. 

In this paper, we introduce the NNV (\textbf{N}eural \textbf{N}etwork \textbf{V}erification) tool, which is a software framework that performs set-based verification for DNNs and learning-enabled CPS, known colloquially as neural network control systems (NNCS) as shown in Figure~\ref{fig:NCS}\footnote{The source code for NNV is publicly available: \url{https://github.com/verivital/nnv/}. A CodeOcean capsule is also available: \url{https://doi.org/10.24433/CO.1314285.v1}, which will be updated with a new DOI and the latest reproducibility results if accepted. The latest version of the CodeOcean capsule with all aspects described in this paper is available at: \url{https://codeocean.com/capsule/1314285/}, which requires a username (taylor.johnson@uta.edu) and password (cav2020ae) to access. This account has read-only permission, so to rerun the results shown in the capsule, you can select Capsule then Duplicate from the menu bar, which will clone the capsule to allow rerunning and editing if desired. Detailed instructions for the artifact evaluation are available at: \url{https://github.com/verivital/run_nnv_comparison/blob/cav2020/README_AE.md}}.
NNV provides a set of reachability algorithms that can compute both the exact and over-approximate reachable sets of DNNs and NNCSs using a variety of set representations such as polyhedra \cite{tran2019parallel, xiang2018specification, xiang2017reachable, xiang2018reachable, xiang2018output}, star sets \cite{tran2019fm, tran2019emsoft,tran2019safe, manzanas2019arch}, zonotopes \cite{singh2018fast}, and abstract domain representations\cite{singh2019abstract}.
The reachable set obtained from NNV contains all possible states of a DNN from bounded input sets or of a NNCS from sets of initial states of a plant model.
NNV declares a DNN or a NNCS to be safe if, and only if, their reachable sets do not violate safety properties (i.e., have a non-empty intersection with any state satisfying the negation of the safety property).
If a safety property is violated, NNV can construct a complete set of counter-examples demonstrating the set of all possible unsafe initial inputs and states by using the star-based exact reachability algorithm \cite{tran2019fm, tran2019emsoft}.
To speed up computation, NNV uses parallel computing, as the majority of the reachability algorithms in NNV are more efficient when executed on multi-core platforms and clusters.

\begin{figure}[t]%
\vspace{-1em}%
  \centering%
    \includegraphics[width=\textwidth]{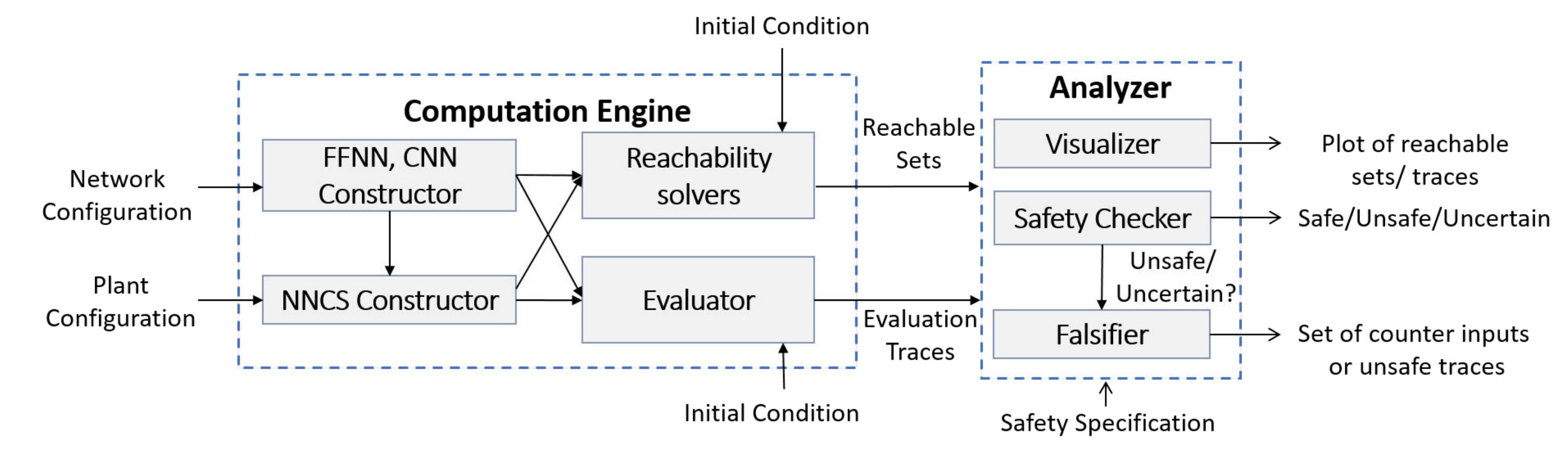}%
		\vspace{-1em}%
   \caption{An overview of NNV and its major modules and components.}%
	\vspace{-1em}%
   \label{fig:NNV}%
	\vspace{-1em}%
\end{figure}%

NNV has been successfully applied to safety verification and robustness analysis of several real-world DNNs, primarily feedforward neural networks (FFNNs) and convolutional neural networks (CNNs), as well as learning-enabled CPS.
To highlight NNV's capabilities, we present brief experimental results from two case studies.
The first compares methods for safety verification of the ACAS Xu networks \cite{julian2016policy}, and the second presents safety verification of a learning-based adaptive cruise control (ACC) system.
%

\section{Overview and Features}%
\vspace{-0.5em}%
NNV is an object-oriented toolbox written in Matlab, which was chosen in part due to the prevalence of Matlab/Simulink in the design of CPS.
It uses the MPT toolbox \cite{kvasnica2004multi} for polytope-based reachability analysis and visualization \cite{tran2019parallel}, and makes use of CORA \cite{althoff2015introduction} for zonotope-based reachability analysis of nonlinear plant models \cite{tran2019emsoft}.
NNV also utilizes the Neural Network Model Transformation Tool (NNMT) for transforming neural network models from Keras and Tensorflow into Matlab using the Open Neural Network Exchange (ONNX) format, and the Hybrid Systems Model Transformation and Translation tool (HyST)~\cite{bak2015hyst} for plant configuration.

The NNV toolbox contains two main modules: a \emph{computation engine} and an \emph{analyzer}, shown in Figure \ref{fig:NNV}.
The computation engine module consists of four subcomponents: 1) the \emph{FFNN constructor}, 2) the \emph{NNCS constructor}, 3) \emph{the reachability solvers}, and 4) \emph{the evaluator}.
The FFNN constructor takes a network configuration file as an input and generates a FFNN object.
The NNCS constructor takes the FFNN object and the plant configuration, which describes the dynamics of a system, as inputs and then creates an NNCS object.
Depending on the application, either the FFNN (or NNCS) object will be fed into a reachability solver to compute the reachable set of the FFNN (or NNCS) from a given initial set of states.
Then, the obtained reachable set will be passed to the analyzer module.
The analyzer module consists of three subcomponents: 1) a \emph{visualizer}, 2) a \emph{safety checker}, and 3) a \emph{falsifier}.
The visualizer can be called to plot the obtained reachable set.
Given a safety specification, the safety checker can reason about the safety of the FFNN or NNCS with respect to the specification.
When an exact (sound and complete) reachability solver is used, such as the star-based solver, the safety checker can return either "safe," or "unsafe" along with a set of counterexamples.
When an over-approximate (sound) reachability solver is used, such as the zonotope-based scheme or the approximate star-based solvers, the safety checker can return either "safe" or "\emph{uncertain}" (unknown).
In this case, the falsifier automatically calls the evaluator to generate simulation traces to find a counterexample.
If the falsifier can find a counterexample, then NNV returns unsafe.
Otherwise, it returns unknown.
A summary of NNV's major features is given in Table \ref{tab:nnv-features}.
\begin{table}[]%
  \centering%
  \resizebox{0.875\linewidth}{!}{
\setlength{\arrayrulewidth}{.15em}
    \begin{tabular}{lll}
\hline
\textbf{Feature}             & \textbf{Exact Analysis} & \textbf{Over-approximate Analysis} \\ \hline
Components                          & \href{https://github.com/verivital/nnv/tree/master/code/nnv/engine/nn/fnn}{FFNN}, \href{https://github.com/verivital/nnv/tree/master/code/nnv/engine/nn/cnn}{CNN}, \href{https://github.com/verivital/nnv/tree/master/code/nnv/engine/nncs}{NNCS}         & \href{https://github.com/verivital/nnv/tree/master/code/nnv/engine/nn/fnn}{FFNN}, \href{https://github.com/verivital/nnv/tree/master/code/nnv/engine/nn/cnn}{CNN}, \href{https://github.com/verivital/nnv/tree/master/code/nnv/engine/nncs}{NNCS}                         \\
Plant dynamics (for NNCS)           & \href{https://github.com/verivital/nnv/blob/master/code/nnv/engine/nncs/LinearODE.m}{Linear ODE}              & \href{https://github.com/verivital/nnv/blob/master/code/nnv/engine/nncs/LinearODE.m}{Linear ODE}, \href{https://github.com/verivital/nnv/blob/master/code/nnv/engine/nncs/NonLinearODE.m}{Nonlinear ODE}          \\
Discrete/Continuous (for NNCS)      & Discrete Time           & Discrete Time, Continuous Time              \\
Activation functions                & \href{https://github.com/verivital/nnv/blob/master/code/nnv/engine/nn/fnn/ReLU.m}{ReLU}, \href{https://github.com/verivital/nnv/blob/master/code/nnv/engine/nn/fnn/SatLin.m}{Satlin}            & \href{https://github.com/verivital/nnv/blob/master/code/nnv/engine/nn/fnn/ReLU.m}{ReLU}, \href{https://github.com/verivital/nnv/blob/master/code/nnv/engine/nn/fnn/SatLin.m}{Satlin}, \href{https://github.com/verivital/nnv/blob/master/code/nnv/engine/nn/fnn/LogSig.m}{Sigmoid}, \href{https://github.com/verivital/nnv/blob/master/code/nnv/engine/nn/fnn/TanSig.m}{Tanh}        \\
CNN Layers													& \href{https://github.com/verivital/nnv/blob/master/code/nnv/engine/nn/cnn/MaxPooling2DLayer.m}{MaxPool}, \href{https://github.com/verivital/nnv/blob/master/code/nnv/engine/nn/cnn/Conv2DLayer.m}{Conv}, \href{https://github.com/verivital/nnv/blob/master/code/nnv/engine/nn/cnn/BatchNormalizationLayer.m}{BN}, \href{https://github.com/verivital/nnv/blob/master/code/nnv/engine/nn/cnn/AveragePooling2DLayer.m}{AvgPool}, \href{https://github.com/verivital/nnv/blob/master/code/nnv/engine/nn/cnn/FullyConnectedLayer.m}{FC} & \href{https://github.com/verivital/nnv/blob/master/code/nnv/engine/nn/cnn/MaxPooling2DLayer.m}{MaxPool}, \href{https://github.com/verivital/nnv/blob/master/code/nnv/engine/nn/cnn/Conv2DLayer.m}{Conv}, \href{https://github.com/verivital/nnv/blob/master/code/nnv/engine/nn/cnn/BatchNormalizationLayer.m}{BN}, \href{https://github.com/verivital/nnv/blob/master/code/nnv/engine/nn/cnn/AveragePooling2DLayer.m}{AvgPool}, \href{https://github.com/verivital/nnv/blob/master/code/nnv/engine/nn/cnn/FullyConnectedLayer.m}{FC} \\
Reachability methods                & \href{https://github.com/verivital/nnv/blob/master/code/nnv/engine/set/Star.m}{Star}, Polyhedron, \href{https://github.com/verivital/nnv/blob/master/code/nnv/engine/set/ImageStar.m}{ImageStar}        & \href{https://github.com/verivital/nnv/blob/master/code/nnv/engine/set/Star.m}{Star}, \href{https://github.com/verivital/nnv/blob/master/code/nnv/engine/set/Zono.m}{Zonotope}, Abstract-domain, \href{https://github.com/verivital/nnv/blob/master/code/nnv/engine/set/ImageStar.m}{ImageStar}     \\
Reachable set/Flow-pipe Visualization                         & Yes                     & Yes                                \\
Parallel computing                  & Yes                     & Partially supported                \\
Safety verification                 & Yes                     & Yes                                \\
Falsification                       & Yes                     & Yes                                 \\
Robustness verification (for FFNN/CNN)  & Yes                     & Yes                                \\
Counterexample generation          & Yes                       & Yes                                 \\
%
\hline
\end{tabular}}%
\caption{Overview of major features available in NNV. Links refer to relevant files/classes in the NNV codebase. BN refers to batch normalization layers, FC to fully-connected layers, AvgPool to average pooling layers, Conv to convolutional layers, and MaxPool to max pooling layers.}%
\label{tab:nnv-features}%
\vspace{-2em}%
\end{table}%

\section{Set Representations and Reachability Algorithms}
\begin{figure}[t]%
\vspace{-1em}%
  \centering%
    \includegraphics[width=0.7\textwidth]{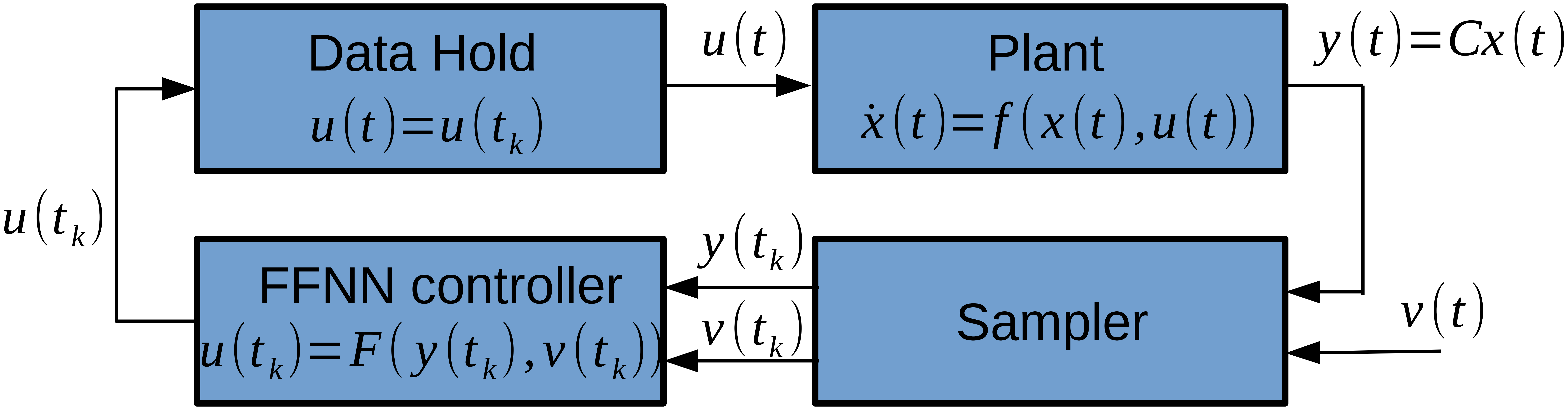}%
		\vspace{-1em}%
   \caption{Architecture of a typical neural network control system (NNCS).}%
	\vspace{-1em}
   \label{fig:NCS}%
	\vspace{-1.5em}%
 \end{figure}%
NNV implements a set of reachability algorithms for \emph{sequential} FFNNs and CNNs, as well as NNCS with FFNN controllers as shown in Figure \ref{fig:NCS}.
The reachable set of a sequential FFNN is computed layer-by-layer.
The output reachable set of a layer is the input set of the next layer in the network.

\subsection{Polyhedron \cite{tran2019parallel} } 
The polyhedron reachability algorithm computes the exact polyhedron reachable set of a FFNN with ReLU activation functions. The exact reachability computation of layer $L$ in a FFNN is done as follows. First, we construct the affine mapping $\bar{I}$ of the input polyhedron set $I$, using the weight matrix $W$ and the bias vector $b$, i.e., $\bar{I} = W\times I + b$. Then, the exact reachable set of the layer $R_L$ is constructed by executing a sequence of stepReLU operations, i.e., $R_{L} = stepReLU_n(stepReLU_{n-1}(\cdots (stepReLU_1(\bar{I}))))$. Since a $stepReLU$ operation can split a polyhedron into two new polyhedra, the exact reachable set of a layer in a FFNN is usually a union of polyhedra. The polyhedron reachability algorithm is computationally expensive because computing affine mappings with polyhedra is costly. Additionally, when computing the reachable set, the polyhedron approach extensively uses the expensive conversion between the H-representation and the V-representation. These are the main drawbacks that limit the scalability of the polyhedron approach. Despite that, we extend the polyhedron reachability algorithm for NNCSs with FFNN controllers. However, the propagation of polyhedra in NNCS may lead to a large degree of conservativeness in the computed reachable set \cite{tran2019emsoft}.

\subsection{Star Set \cite{tran2019fm,tran2019emsoft} (\href{https://github.com/verivital/nnv/blob/master/code/nnv/engine/set/Star.m}{code})}

The star set is an efficient set representation for simulation-based verification of large linear systems \cite{bak2017simulation, bak2019numerical, tran2019formats} where the superposition property of a linear system can be exploited in the analysis. It has been shown in \cite{tran2019fm} that the star set is also suitable for reachability analysis of FFNNs. In contrast to polyhedra, the affine mapping and intersection with a half space of a star set is more easily computed. NNV implements an enhanced version of the exact and over-approximate reachability algorithms for FFNNs proposed in \cite{tran2019fm} by minimizing the number of LP optimization problems that need to be solved in the computation. The exact algorithm that makes use of star sets is similar to the polyhedron method that makes use of $stepReLU$ operations. However, it is much faster and more scalable than the polyhedron method because of the advantage that star sets have in affine mapping and intersection. The approximate algorithm obtains an over-approximation of the exact reachable set by approximating the exact reachable set after applying an activation function, e.g., ReLU, Tanh, Sigmoid. We refer readers to \cite{tran2019fm} for a detailed discussion of star-set reachability algorithms for FFNNs.

We note that NNV implements enhanced versions of earlier star-based reachability algorithms \cite{tran2019fm}.
Particularly, we minimize the number of linear programming (LP) optimization problems that must be solved in order to construct the reachable set of a FFNN by quickly estimating the ranges of all of the states in the star set using only the ranges of the predicate variables. 
Additionally, the extensions of the star reachability algorithms to NNCS with linear plant models can eliminate the explosion of conservativeness in the polyhedron method \cite{tran2019emsoft, tran2019safe}. The reason behind this is that in star sets, the relationship between the plant state variables and the control inputs is preserved in the computation since they are defined by a unique set of predicate variables. We refer readers to \cite{tran2019emsoft, tran2019safe} for  a detailed discussion of the extensions of the star-based reachability algorithms for NNCSs with linear/nonlinear plant models.

\subsection{Zonotope \cite{singh2018fast} (\href{https://github.com/verivital/nnv/blob/master/code/nnv/engine/set/Zono.m}{code})}

NNV implements the zonotope reachability algorithms proposed in \cite{singh2018fast} for FFNNs. Similar to the over-approximate algorithm using star sets, the zonotope algorithm computes an over-approximation of the exact reachable set of a FFNN. Although the zonotope reachability algorithm is very fast and scalable, it produces a very conservative reachable set in comparison to the star set method as shown in \cite{tran2019fm}. Consequently, zonotope-based reachability algorithms are usually only more efficient for very small input sets. As an example it can be more suitable for robustness certification.

\subsection{Abstract Domain \cite{singh2019abstract}}

NNV implements the abstract domain reachability algorithm proposed in \cite{singh2019abstract} for FFNNs. NNV's abstract domain reachability algorithm specifies an abstract domain as a star set and uses a ``back-tracking'' approach to estimate the \emph{over-approximate ranges} of the states. The abstract domain is more conservative than the star set method.

\subsection{ImageStar Set (\href{https://github.com/verivital/nnv/blob/master/code/nnv/engine/set/ImageStar.m}{code})}
NNV recently introduced a new set representation called the ImageStar for use in the verification of deep convolutional neural networks (CNNs). Briefly, the ImageStar is a generalization of the star set where the anchor and generator vectors are replaced by multi-channel images. The ImageStar is efficient in the analysis of convolutional layers, average pooling layers, and fully connected layers, whereas max pooling layers and ReLU layers consume most of computation time in reachability analysis of CNNs. NNV implements exact and over-approximate reachability algorithms using the ImageStar for serial CNNs. Since the ImageStar method has not been published yet, we defer its evaluation in our experimental evaluation. In short, using the ImageStar, we can analyze the robustness under adversarial attacks of the real-world VGG16 and VGG19 deep perception networks \cite{simonyan2014very} that consist of $>100$ million parameters.

\section{Evaluation}
The experiments presented in this section were performed on a desktop with the following configuration: Intel Core i7-6700 CPU $@$ 3.4GHz 8 core Processor, 64 GB Memory, and 64-bit Ubuntu 16.04.3 LTS OS.
\subsection{Safety verification of ACAS Xu networks}
We evaluate NNV in comparison to Reluplex \cite{katz2017reluplex}, Marabou \cite{katz2019marabou}, and  ReluVal \cite{shiqi2018reluval}, by considering the verification of safety property $\phi_3$,  and $\phi_4$ of the ACAS Xu neural networks \cite{julian2016policy} for all $45$ networks.\footnote{We omitted properties $\phi_1$ and $\phi_2$ for space and due to their long runtimes, but they can be reproduced in the artifact evaluation if desired.}
%
%
%
All the experiments were done using $4$ cores for the computation.
The verification results are summarized in Table \ref{tab:acasxu} where (SAT) denotes that the networks are safe,  (UNSAT) denotes unsafe, and (UNK) denotes unknown.
We note that (UNK) may occur due to the conservativeness of the reachability analysis scheme.
Detailed verification results are presented in the appendix.
For a fast comparison with other tools, we also tested a subset input of Property 1-4 on all the 45 networks.
The results are also shown in the appendix.
We note that the polyhedron method \cite{tran2019parallel} achieves a timeout on most of networks, and therefore, we neglect this method in the comparison.

\begin{table}[]
\centering
 \resizebox{0.8\linewidth}{!}{
\renewcommand{\arraystretch}{1}
\renewcommand{\tabcolsep}{1.0mm}
\begin{tabular}{l|ccccccc}
\hline
\multirow{2}{*}{\textbf{ACAS XU $\phi_3$}} &\multirow{2}{*}{\textbf{SAT}} & \multirow{2}{*}{\textbf{UNSAT}} & \multirow{2}{*}{\textbf{UNK}} & \multicolumn{3}{l}{\textbf{TIMEOUT}}      & \multirow{2}{*}{\textbf{TIME(s)}}  \\ \cline{5-7}
                  &                     &                        &                          & \textbf{1h} & \textbf{2h} &{\textbf{10h}} &                     \\ \hline 
Reluplex          & 3                   & 42                      & 0                        & 2  & 0  & 0                              & 28454                         \\ 
Marabou           & 3                   & 42                      & 0                        & 1  & 0  & 0                              & 19466                       \\ 
Marabou DnC       & 3                   & 42                      & 0                        & 3  & 3  & 1                              & 111880                      \\ 
ReluVal           & 3                   & 42                      & 0                        & 0  & 0  & 0                              & 416                         \\ 
Zonotope          & 0                    & 2                      & 43                       & 0  & 0  & 0                               & 3                         \\ 
Abstract Domain   & 0                   & 10                      & 35                       & 0  & 0  & 0                               & 72                       \\ 
NNV Exact Star    & 3                   & 42                      & 0                        & 0  & 0  & 0                              & 1371                       \\ 
NNV Appr. Star   & 0                   & 29                      & 16                       & 0  & 0  & 0                               & 52                        \\ \hline \hline 
\multirow{1}{*}{\textbf{ACAS XU $\phi_4$}} & \\\hline 
Reluplex          & 3                   & 42                      & 0                        & 0  & 0  & 0                               & 11880                    \\  
Marabou           & 3                   & 42                      & 0                        & 0  & 0  & 0                               & 8470                    \\  
Marabou DnC       & 3                   & 42                      & 0                        & 2  & 2  & 0                               & 25110                   \\  
ReluVal           & 3                   & 42                      & 0                        & 0  & 0  & 0                              & 27                     \\  
Zonotope          & 0                   & 1                      & 44                       & 0  & 0  & 0                               & 5                    \\  
Abstract Domain   & 0                   & 0                      & 45                       & 0  & 0  & 0                               & 7                      \\  
NNV Exact Star    & 3                   & 42                      & 0                        & 0  & 0  & 0                               & 470                   \\  
NNV Appr. Star    & 0                   & 32                      & 13                        & 0  & 0  & 0                               & 19                     \\ \hline 
\end{tabular}}
\caption{Verification results of ACAS Xu networks.}
\label{tab:acasxu}
\vspace{-2em}
\end{table}

\textbf{Verification time.}
%
%
%
%
%
%
For property $\phi_3$, our exact-star method is about $20.7\times$ faster than Reluplex, $14.2\times$ faster than Marabou, $81.6\times$ faster than Marabou-DnC (i.e., divide and conquer method). The approximate star method is $547\times$ faster than Reluplex, $374\times$  faster than Marabou, $2151\times$ faster than Marabou-DnC, and $8\times$ faster than ReluVal.
For property $\phi_4$, our exact-star method is $25.3\times$ faster than Reluplex, $18.0\times$ faster than Marabou, $53.4\times$ faster than Marabou-DnC, while the approximate star method is $625\times$ faster than Reluplex, $445\times$ faster than Marabou, $1321\times$ faster than Marabou-DnC.
%
%

%
%
%
%

\textbf{Conservativeness.} The approximate star method is much less conservative than the zonotope and abstract domain methods.
This is illustrated since it can verify more networks than the zonotope and abstract domain methods, and is because it obtains a tighter over-approximate reachable set.
For property $\phi_3$, the zonotope and abstract domain methods can prove the safety of $2/45$ networks, ($4.44\%$) and $19/45$ networks, ($42.22\%$) respectively, while our approximate star method can prove the safety of $29/45$ networks, ($64.4\%$ ).
For property $\phi_4$, the zonotope and abstract domain method can prove the safety of $1/45$ networks, ($2.22\%$) and $0/45$ networks, ($0.00\%$) respectively while the approximate star method can prove the safety of $32/45$, ($71.11\%$).

\subsection{Safety Verification of Adaptive Cruise Control System}
\begin{figure*}[t!]%
\vspace{-1em}%
  \centering%
    \includegraphics[width=\textwidth]{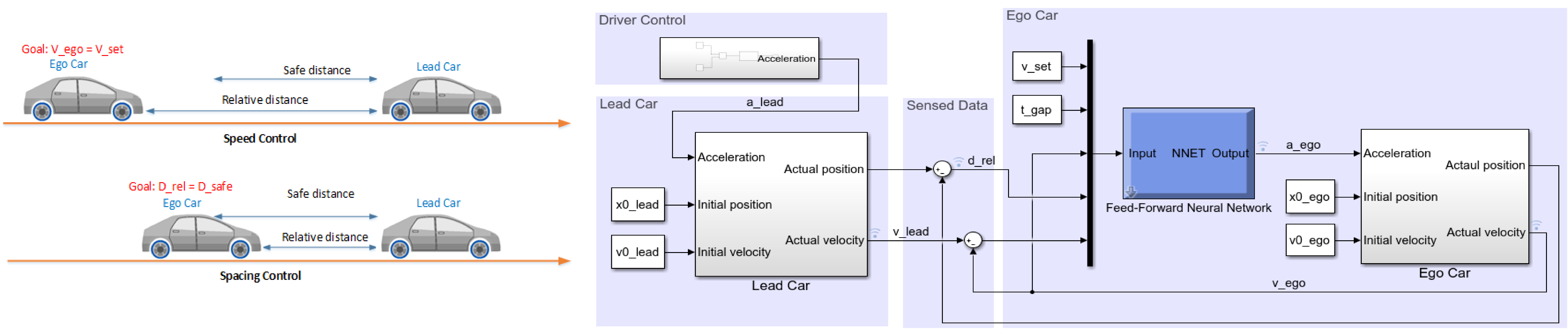}%
   \caption{Learning-based Adaptive Cruise Control System~\cite{matlab2019acc}.}%
	\vspace{-10pt}%
   \label{fig:ACC}%
	\vspace{-5pt}%
 \end{figure*}

To illustrate how NNV can be used to verify/falsify safety properties of learning-enabled CPS, we analyze a learning-based ACC system depicted in Figure \ref{fig:ACC}, in which the ego vehicle has a radar sensor to measure the distance to the lead vehicle in the same lane, $D_{rel}$, as well as the relative velocity of the lead vehicle, $V_{rel}$.
The ego vehicle has two control modes.
In speed control mode, it travels at a driver-specified set speed $V_{set} = 30$, and in spacing control mode, it maintains a safe distance from the lead vehicle, $D_{safe}$.
We train a neural network with $5$ layers, $20$ neurons per layer utilizing the ReLU activation function to control the ego vehicle with a control period of $0.1$ seconds. 

We investigate safety of the learning-based ACC system with two types of plant dynamics: 1) a discrete linear plant, and 2) a nonlinear continuous plant governed by the following differential equations:%
\begin{align*}
  &\dot{x}_{lead}(t) = v_{lead}(t),~\dot{v}_{lead}(t) = \gamma_{lead},  & \dot{\gamma}_{lead}(t) = -2 \gamma_{lead}(t) + 2a_{lead} - \mu v^2_{lead}(t), \\
  &\dot{x}_{ego}(t) = v_{ego}(t),~\dot{v}_{ego}(t) = \gamma_{ego},  & \dot{\gamma}_{ego}(t) = -2 \gamma_{ego}(t) + 2a_{ego} - \mu v^2_{ego}(t),
\end{align*}
where $x_{lead} (x_{ego})$, $v_{lead} (v_{ego})$ and $\gamma_{lead} (\gamma_{ego})$ are the position, velocity and acceleration of the lead (ego) vehicle respectively. $a_{lead}(a_{ego})$ is the acceleration control input applied to the lead (ego) vehicle, and $\mu = 0.0001$ is a friction parameter.
To obtain a discrete linear model of the plant, we let $\mu = 0$ and discretize the corresponding linear continuous model using a zero-order hold on the inputs with a sample time of $0.1$ seconds (i.e., the control period). 

\textbf{Verification Problem.} The scenario we are interested in is when the two vehicles are operating at a safe distance between them and the ego vehicle is in speed control mode. In this state the lead vehicle driver suddenly decelerates with $a_{lead} = -5$ to reduce the speed. 
We want to verify if the neural network controller on the ego vehicle will also de-accelerate to maintain a safe distance between the two vehicles.
To guarantee safety, we require that $D_{rel} = x_{lead} - x_{ego} \geq D_{safe} =  D_{default} + T_{gap} \times v_{ego}$ where $T_{gap} = 1.4$ seconds and $D_{default} = 10$. 
Our analysis investigates if the safety requirement holds in the $5$ seconds after the lead vehicle decelerates. We consider the safety of the system under the following initial conditions: $x_{lead}(0) \in [90, 92]$, $v_{lead}(0) \in [20, 30]$, $\gamma_{lead}(0) = \gamma_{ego}(0) = 0$, $v_{ego}(0) \in [30, 30.5]$, $x_{ego} \in [30, 31]$.

\begin{figure}[t!]%
  \centering%
    \includegraphics[width=0.7\textwidth]{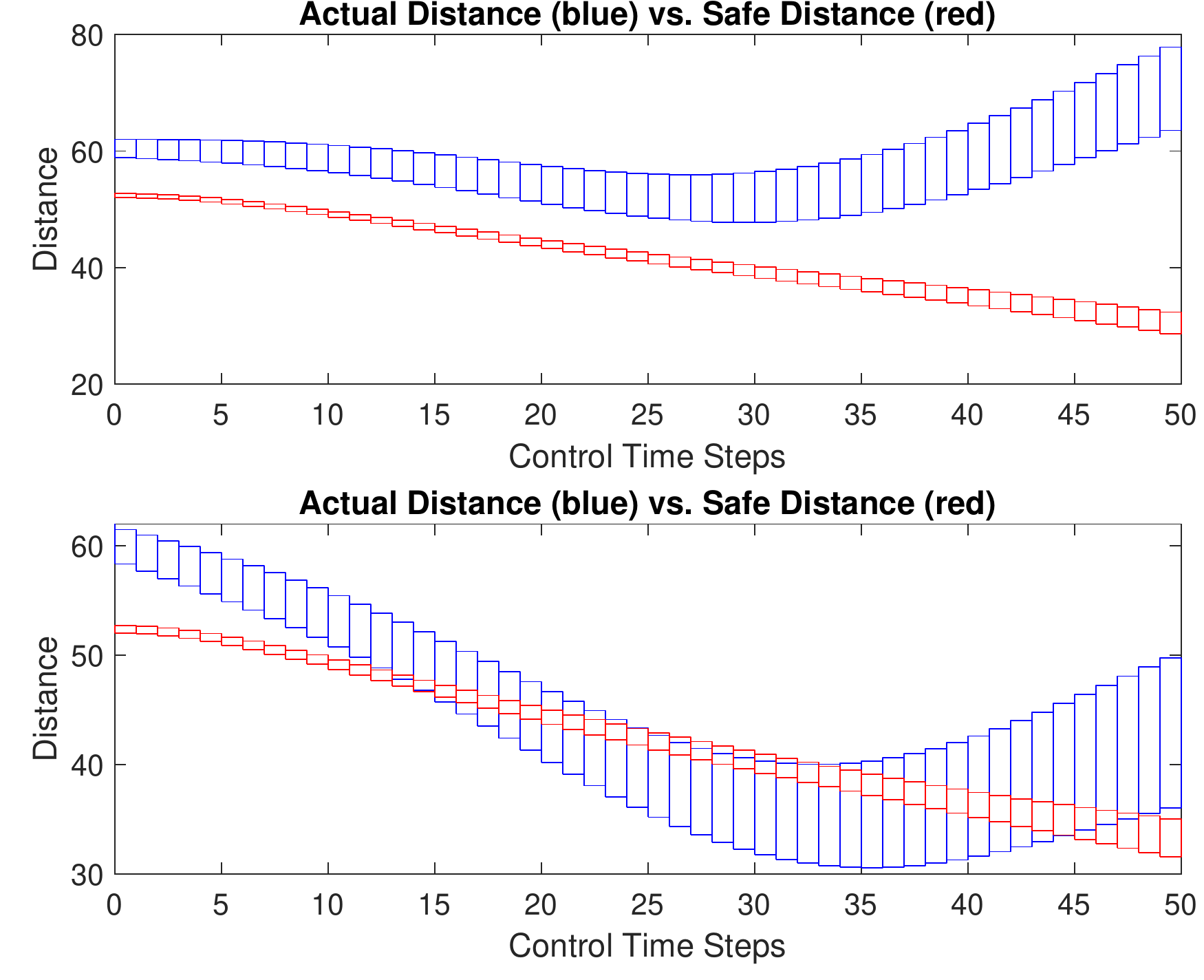}%
		\vspace{-5pt}%
   \caption{Two scenarios of the ACC system. In the first (top) scenario ($v_{lead}(0) \in [29,~30] m/s$), safety is guaranteed, $D_{rel} \geq D_{safe}$. In the second scenario (bottom) ($v_{lead}(0) \in [24,~25] m/s$), safety is violated since $D_{ref} < D_{safe}$ in some control steps.}%
	\vspace{-5pt}%
   \label{fig:ACC-reachSet}%
	\vspace{-1em}%
 \end{figure}

\textbf{Verification results.}
For linear dynamics, NNV can compute both the exact and over-approximate reachable sets of the ACC system in bounded time steps, while for nonlinear dynamics, NNV constructs an over-approximation of the exact reachable sets and uses it for safety verification.
The verification results for linear and nonlinear models using the over-approximate star method are presented in Table \ref{tab:acc}, which shows that, the safety of the ACC system depends on the initial velocity of the lead vehicle.
When the initial velocity of the lead vehicle is smaller than $27 (m/s)$, the ACC system with the discrete plant model is unsafe.
Using the exact star method, NNV can construct a \emph{complete} set of counter-example inputs.
When the over-approximate star method is used, if there is a potential safety violation, NNV simulates the system with $1000$ random inputs from the input set to find counter examples.
If a counterexample is found, the system is \emph{UNSAFE}, otherwise, NNV returns a safety result of \emph{UNKNOWN}.
Figure \ref{fig:ACC-reachSet} visualizes the reachable sets of the relative distance $D_{rel}$ between two vehicles versus the required safe distance $D_{safe}$ over time for two cases of initial velocities of the lead vehicle: $v_{lead}(0) \in [29, 30]$ and $v_{lead}(0) \in [24, 25]$.
We can see that in the first case, $D_{ref} \geq D_{safe}$ for all $50$ time steps stating that the system is safe.
In the second case, $D_{ref} < D_{safe}$ in some control steps which means that the system is unsafe.
NNV supports a \emph{reachLive} method to perform analysis and reachable set visualization on-the-fly to help the user observe the behavior of the system during verification.

The verification results for ACC system with the nonlinear model are all $UNSAFE$, which is surprising.
Since the neural network controller of the ACC system was trained with the linear model, it works quite well for the linear model.
However, when a small friction term is added to the linear model to form a nonlinear model, the neural network controller's performance, in terms of safety, is significantly reduced.
This problem raises an important issue in training neural network controllers using simulation data, and these schemes may not work in real systems since there is always a mismatch between the plant model in the simulation engine and the real system.

\begin{table}[t!]%
    \centering%
   \resizebox{0.5\linewidth}{!}{
\begin{tabular}{lcccl}
\hline
\multirow{2}{*}{\textbf{v\_lead(0)}~~} & \multicolumn{2}{c}{\textbf{Linear Plant}~~~~~~} & \multicolumn{2}{l}{\textbf{Nonlinear Plant}} \\ \cline{2-5}
                                     & $Safety~~~$              & $VT (s)~~~~~~$              & $Safety~~~$                & $VT (s)$               \\ \hline
{[}29, 30{]}                         & SAFE                &      $9.60$               & UNSAFE~~~                & $346.62$                      \\
{[}28, 29{]}                         & SAFE                &      $9.45$           & UNSAFE~~~                		 & $277.50$                     \\
{[}27, 28{]}                         & SAFE                &      $9.82$               & UNSAFE~~~                & $289.70$                     \\
{[}26, 27{]}                         & UNSAFE              &      $17.80$           & UNSAFE~~~                	 & $315.60$                     \\
{[}25, 26{]}                         & UNSAFE              &      $19.24$               & UNSAFE~~~               & $305.56$                     \\
{[}24, 25{]}                         & UNSAFE              &      $18.12$               & UNSAFE~~~               & $372.00$                     \\ \hline
\end{tabular}}%
\caption{Verification results for ACC system with different plant models, where $VT$ is the verification time (in seconds).}%
\label{tab:acc}%
\vspace{-2.25em}%
\end{table}%

%

%
%

\textbf{Verification times.}
As shown in Table~\ref{tab:acc}, the approximate analysis of the ACC system with discrete linear plant model is very fast.
It can be done in $84$ seconds.
We note that NNV also supports exact analysis, which is computationally expensive since it constructs all reachable sets of the system.
Because there are splits in the reachable sets of the neural network controller, the number of star sets in the reachable set of the plant increases quickly over time \cite{tran2019emsoft}.
In contrast, the over-approximate method computes the interval hull of all reachable star sets at each time step.
It maintains a single reachable set of the plant throughout the computation.
Therefore, the over-approximate method is much faster than the exact method.
In terms of plant models, the nonlinear model requires more computation time than the linear one.
As shown in Table~\ref{tab:acc}, the verification for linear model using the over-approximate method is $22.7\times$ faster on average than the verification of the nonlinear model.   




\section{Related Work}

NNV was inspired by many insightful research works in the emerging fields of neural network and machine learning verification.
For the ``open-loop'' verification problem (verification of DNNs), many efficient techniques have been proposed, such as SMT-based methods \cite{pulina2010abstraction,katz2017reluplex,katz2019marabou}, mixed-integer linear programming methods \cite{lomuscio2017approach, kouvaros2018formal,dutta2017output}, set-based methods\cite{xiang2018output,wang2018formal,wang2018efficient,gehr2018ai,singh2018fast,singh2019abstract,anderson2019pldi}, and optimization methods \cite{weng2018towards,zhang2018efficient}.
For the ``closed-loop'' verification problem (NCCS verification), we note that the Verisig approach \cite{ivanov2018verisig} is very efficient for NNCS with nonlinear plants and with Sigmoid and Tanh activation functions.
Additionally, the recent regressive polynomial rule inference approach \cite{souradeep2019} is very fast for the safety verification of NNCS with nonlinear plant models and ReLU activation functions.
The satisfiability modulo convex (SMC) approach \cite{sun2018formal} is also very promising for NNCS with discrete linear plants as it provides both soundness and completeness properties in verification.
ReachNN \cite{huang2019reachnn} is a recent approach that can efficiently control the conservativeness in the reachability analysis of NNCS with nonlinear plants and ReLU, Sigmoid and Tanh activation functions in the controller.
In other learning-enabled systems, falsification and testing-based approaches \cite{dreossi2019cav,tuncali2018simulation, dreossi2017compositional} have shown a significant promise in enhancing the safety of systems where perception components and neural network controllers interact with the physical world.
Finally, there is significant related work in the domain of safe reinforcement learning~\cite{alshiekh2018aaai,verma2018icml,fulton2019tacas,zhu2019pldi} and combining guarantees from NNV with those provided in these methods would be interesting to explore.
\section{Conclusion and Future Work}
We have presented NNV, a toolbox for the verification of DNNs and learning-enabled CPS.
Our tool provides a collection of reachability algorithms that can be used to verify the safety (and robustness) of real-world DNNs as well as learning-enabled CPS, such as the ACC system.
Our method is comparable to existing methods such as Reluplex and Marabou when dealing with the open-loop verification problem.
For closed-loop systems, NNV can compute the exact and over-approximate reachable sets of a NNCS with linear plant models.
For a NNCS with a nonlinear plant, NNV can obtain an over-approximate reachable set and use it to verify the safety, but can also automatically falsify the system to construct/find counterexamples (using exact analysis) or randomized simulations (in over-approximate analysis). 
%



\newpage

\normalsize
\let\oldbibliography\thebibliography
\renewcommand{\thebibliography}[1]{\oldbibliography{#1}
\setlength{\itemsep}{0pt}} 
\bibliographystyle{splncs03}
\bibliography{tran,cav2020}  

\newpage

\appendix
\section{Appendix: Additional Evaluation Details}
\begin{figure*}[b!]
  \centering
    \includegraphics[width=\textwidth]{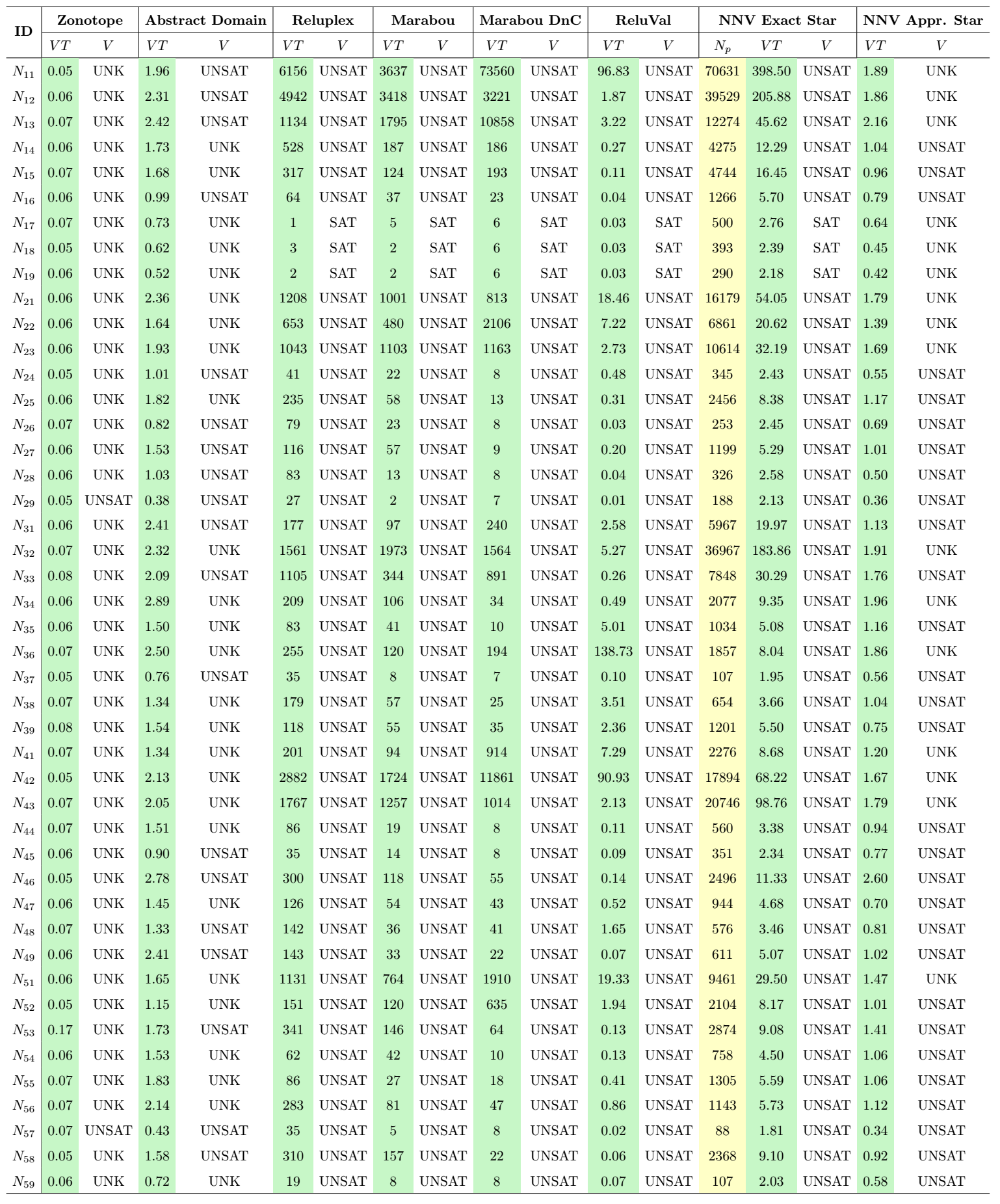}
   \caption{Detailed verification results for $\phi_3$ of ACAS Xu networks.}
	\label{fig:acasxu_phi3}
 \end{figure*}

\begin{figure*}[t!]
  \centering
    \includegraphics[width=\textwidth]{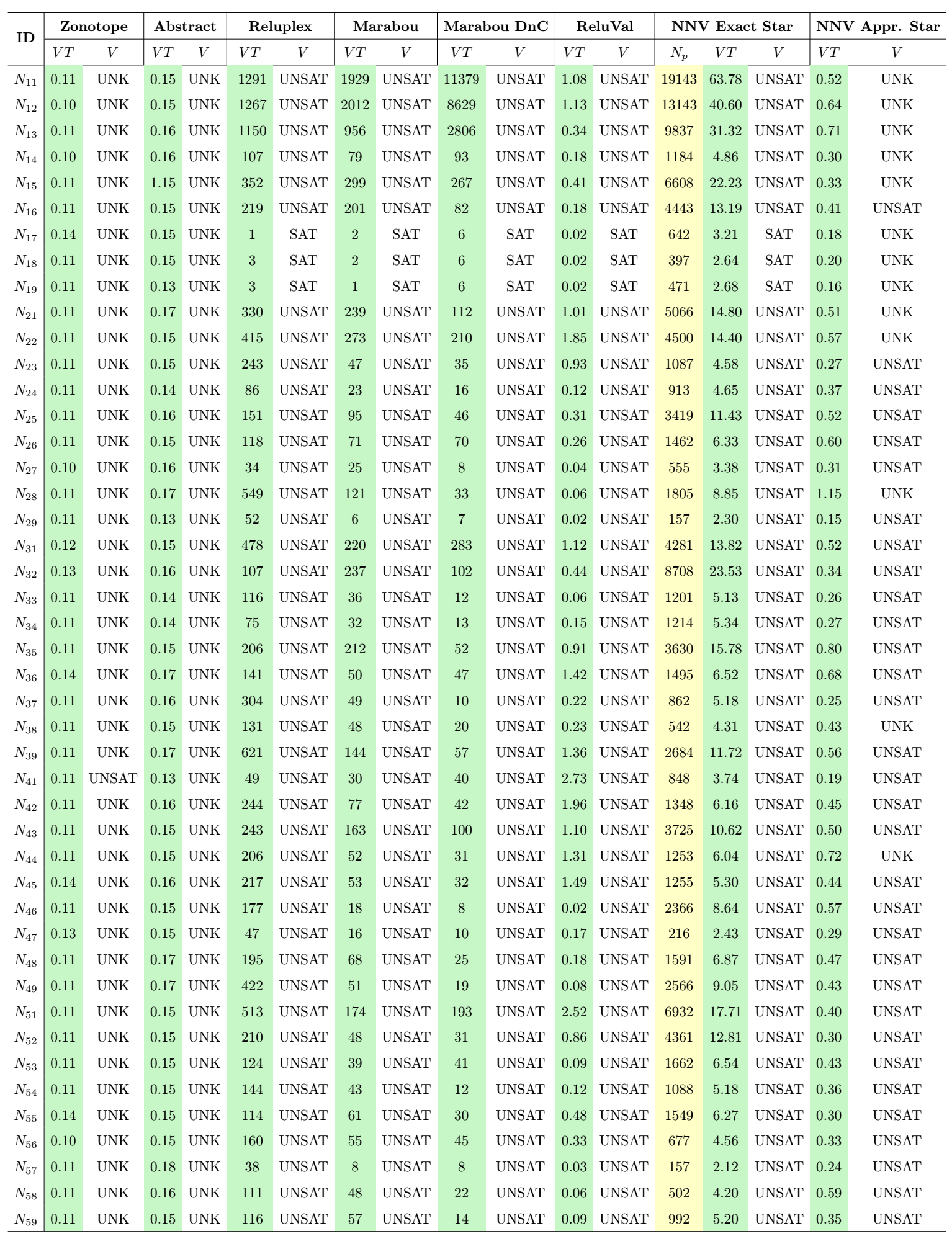}
   \caption{Detailed verification results for $\phi_4$ of ACAS Xu networks.}
	\label{fig:acasxu_phi4}
 \end{figure*}

Figures~\ref{fig:acasxu_phi3} and~\ref{fig:acasxu_phi4} show detailed comparisons between NNV's approximate and exact star methods relative to the zonotope, abstract domain, Reluplex, Marabou, and Marabou divide-and-conquer (DnC) methods, summarized earlier in Table~\ref{tab:acasxu}.

\end{document}